\documentclass[a4paper, 12pt]{article}
\usepackage[english]{babel}
\usepackage[a4paper, inner=2cm, outer=2cm, top=3cm, bottom=3cm]{geometry}
\usepackage[dvipsnames]{xcolor}
\usepackage{mathtools}
\usepackage{pstricks}
\usepackage{caption}
\usepackage{graphicx}
\usepackage{float}
\usepackage{amsmath}
\usepackage{amssymb}
\usepackage{mathcomp}
\usepackage{cancel}
\usepackage{textcomp}
\usepackage{enumitem}
\usepackage{physics}
\usepackage{romannum}
\usepackage[doublespacing]{setspace}
\usepackage{subcaption}
\usepackage{hyperref}
\usepackage{xcolor}
\usepackage{booktabs}
\usepackage{cite}
\usepackage{bm}
\usepackage{authblk}
\usepackage{gensymb}

\hypersetup{
  colorlinks   = true,
  urlcolor     = Blue, 
  linkcolor    = Magenta,
  citecolor    = Magenta
}

\newcommand{\Ddv}[2]{\frac{\mathrm{D} #1}{\mathrm{D} #2}}

\title{\textbf{On the Dynamics of Extended Bodies and the GravoThermo Memory Effect}}
\author[1]{Raihaneh Moti\thanks{\href{mailto:r.moti@ipm.ir}{\ r.moti@ipm.ir}}}
\author[2]{Ali Shojai\thanks{\href{mailto:ashojai@ut.ac.ir}{\ ashojai@ut.ac.ir}  (Corresponding Author)}}
\affil[1]{{\small\textit{School of Astronomy, Institute for Research in Fundamental Sciences (IPM), \hspace{4cm} Tehran, Iran P.O. Box 19395-5531}} }
\affil[2]{{\small\textit{Department of Physics, University of Tehran, North Karegar St., Tehran, Iran}}}

\date{}

\date{}\begin{document}
\maketitle
\pagenumbering{arabic}

\begin{abstract}
The non-linearity of the theory of gravity induces a hysteresis effect in both the systems interacting with gravity and in the gravitational field. The effect is usually referred to as the memory effect. In this paper, we explore this phenomenon in the context of the dynamics of extended rotating objects in the presence of a gravitational wave pulse. Then, we consider an ensemble of such objects and show that  
a redistribution of spin orientations takes place as a gravitational wave passes through.
We examine how this redistribution, changes the partition function and the thermodynamic quantities including entropy and energy of the system. The result shows that some information about the source of the gravitation wave is encoded in the thermodynamics of the system.
\end{abstract}

%%%%%%%%%%%%%%%%%%%%%%%%%%%%%%%%%%%% Section I
\section{Introduction}
\label{Sec:Intro}
The dynamics of physical systems are governed by differential equations, which are either linear or non-linear. Solutions to linear systems, aside from being easier to analyze, obey the superposition principle. Furthermore, the trajectory is often time--reversible in many such systems.

In contrast, solutions to non-linear dynamical equations do not satisfy the superposition principle. Consequently, these systems may exhibit hysteresis-like behavior, giving rise to memory effects in their interactions with the environment. This allows the environment to remember past interactions with the system, making it possible to distinguish between its states before and after the interaction.

Interaction with the solutions of non-linear Einstein equation offers one of the most compelling settings to study the memory effect \cite{Zeldovich:1974gvh, Braginsky:1987kwo, Christodoulou:1991cr, Blanchet:1992br, Thorne:1992sdb, Divakarla:2021xrd, Grishchuk:1989qa, Favata:2009ii}. Even in the weak field regime, where linear approximations are typically employed, the underlying non-linearity (in the interaction) manifests as gravitational wave memory \cite{Favata:2008ti, Favata:2008yd, Favata:2010zu, Nichols:2017rqr}. While much of the literature focuses on quantum aspects of the memory \cite{Strominger:2014, Pasterski:2015tva, Zhang:2017geq, Pate:2017, Luca:2025, Caldwell:2025}, here we are more interested in the classical origin of this phenomenon, rooted in the non-linearity of Einstein’s equation.

The strength of this less conventional viewpoint on the memory effect lies in its link to the emergence of an arrow of evolution \cite{Moti:2024a, Moti:2024b}. This approach opens the possibility of a statistical analysis of the system with the memory effect serving as a key evolving parameter.

As a toy model,\cite{Moti:2024a} it is shown that the distribution of an ensemble of spinning point particles will be altered permanently after interacting with the memory sector of a gravitational wave. This effect can be observed through changes in the partition function of the system, with memory parameters clearly emerging in these alterations. The same model is also examined in the strong field regime using kinetic theory in curved spacetime, yielding consistent results\cite{Moti:2024b}.

In this paper, our aim is to explore the effect for the case of extended rotating objects in curved spacetime. It is a well known fact that unlike the point-like particles, such objects do not follow geodesics. Instead, their energy--momentum tensor couples to the spacetime curvature through dynamical equations. This makes it particularly interesting to investigate how the memory effect of the gravitational field influences their motion. We explore this using the Mathisson--Papapetrou formalism. Subsequently, we consider an ensemble of those rotating bodies within a perturbed spacetime, and examine how their partition function will be modified after interacting with a gravitational wave.

In the following sections, we first briefly review the dynamics of an extended rotating object in curved spacetime. In Section \ref{Sec:Pert.}, we consider the perturbation to the flat spacetime induced by a gravitational wave pulse and analyze the solutions of the resulting dynamical equations in detail and investigate the memory effect of body. Next in Section \ref{Sec:NonlinearMemory} nonlinear memory, i.e. the memory of the wave is investigated. Section \ref{Sec:Ensemble} focuses on studying the partition function of an ensemble of those objects. Finally, in Section \ref{Sec:Con.}, we conclude by comparing our results with those obtained in previous studies.

%%%%%%%%%%%%%%%%%%%%%%%%%%%%%%%%%%%% Section II
\section{Mathisson--Papapetrou Equations}
\label{Sec:MpD Eq}
In the general theory of relativity, a particle follows a geodesic curve, unless its physical size becomes comparable to the characteristic length scale of the curved spacetime. When the assumption of having a point particle breaks down, the more fundamental approach based on the covariant conservation of the energy--momentum tensor $T_{\mu\nu}$, 
\begin{equation}
\nabla_{\mu}T^{\mu\nu}=0
\end{equation}
 has to be used which holds regardless of the size of the object. The difference is in fact inherent in the tidal forces interacting locally with $T_{\mu\nu}$  and influencing both the dynamics of the extended body and the energy--momentum tensor. 

Implementation of such a model is practically difficult. A possible workaround is to employ multipole expanded $T_{\mu\nu}$ in the conservation equation, rather than working with the full tensor directly. This often simplifies the analysis by reducing the complexity of equations while preserving the essential features of the dynamics. Based on the foundational work of Mathisson \cite{Mathisson:1937}, Papapetrou \cite{Papapetrou:1951, Corinaldesi:1951}, and Dixon \cite{Dixon:1970}, the general motion of an extended body in curved spacetime is then described through the multipole moment integrals
\begin{equation}
\int T^{\mu\nu} \sqrt{-g} \dd[3]{x}, \quad \int \delta x^{\rho}T^{\mu\nu} \sqrt{-g} \dd[3]{x}, \quad \int \delta x^{\rho}\delta x^{\sigma}T^{\mu\nu} \sqrt{-g} \dd[3]{x}, \quad \ldots
\end{equation}
where $g$ denotes the determinant of the spacetime metric $g_{\mu\nu}$, and the integration is performed over a 3D spatial hypersurface at constant $t$. The $\delta x^{\mu} \equiv x^{\mu} - X^{\mu}$ describes points about the worldline of a specified point (with coordinates $X^{\mu}$) in the object.

According to this, a point particle, which may be called a monopole particle, is defined as an object for which at least one of the integrals of zeroth moment $\int T^{\mu\nu} \sqrt{-g} \dd[3]{x}$ is non-zero, while all higher order moments vanish. A monopole--dipole object, on the other hand, is the one with at least one additional non-vanishing dipole moment integral  $\int \delta x^{\rho}T^{\mu\nu} \sqrt{-g} \dd[3]{x}$ , and vanishing higher moments, and so on.

In the rest frame of the object, the first and second moments, i.e.
\begin{align}
& P_{\nu} = \int T^{0}_{\nu} \sqrt{-g} \dd[3]{x} \ , \label{1st-Moment} \\
& S^{\mu\nu} = \int \left[ \delta x^{\mu} T^{0\nu}-\delta x^{\nu} T^{0\mu} \right] \sqrt{-g} \dd[3]{x} \ , \label{2nd-Moment}
\end{align}
are nothing but the linear momentum and the angular momentum about the point $X^{\mu}$, respectively.

By applying this expansion to the conservation equation of the energy--momentum tensor, a set of dynamical equations corresponding to each moment is obtained where up to the dipole moment is 
\begin{align}
& \Ddv{P_\nu}{\tau} + \dfrac{1}{2} S^{\alpha\mu} R_{\alpha\mu\nu\beta} U^{\beta} = 0 \ , \\
& \Ddv{S^{\mu\nu}}{\tau} + U^{\mu}P^{\nu}-U^{\nu}P^{\mu} = 0 \ .
\end{align}
Here,  $U^{\mu}$ denotes the four-velocity of the reference point worldline $X^{\mu}$ and $\tau$ is an affine parameter along the momentum four-vector $P^{\mu}$. 
The six independent components of $S^{\mu\nu}$, the four components of  $P^{\mu}$, and the three independent components of $U^{\mu}$ (in total thirteen parameters) must satisfy these ten equations which are known as Mathisson--Papapetrou equations.

The apparent redundancy arises from neglecting higher order moments and can be resolved by imposing additional physical conditions such as those derived from Lorentz invariance. The rest frame condition in special relativity is given by $S^{0i}=0$. This condition can be extended to a covariant form as $P^{\mu}S_{\mu\nu} = 0$. Together with the conserved charges $m^2 = P_{\mu} P^{\mu}$ and $s^2 = \frac{1}{2}S_{\mu\nu}S^{\mu\nu}$,  it is not difficult to show that the complete set of differential equations for an extended body is 
\begin{align}
& \Ddv{P_\nu}{\tau} + \dfrac{1}{2} S^{\alpha\mu} R_{\alpha\mu\nu\beta} U^{\beta} = 0 \ , \label{MP-Momentum}\\
& \Ddv{S_{\alpha\mu}}{\tau} + \dfrac{1}{m^2} \left(S_{\alpha\beta} P_{\mu} \Ddv{P^\beta}{\tau} + S_{\beta\mu} P_{\alpha} \Ddv{P^{\beta}}{\tau}    \right) = 0  \ , \label{MP-Spin} \\
& U^{\mu} = \dfrac{\bar{Z}}{m^2} \left( P^{\mu} +\dfrac{1}{2} ~ \dfrac{S^{\mu\nu} R_{\nu\sigma\alpha\beta}P^{\sigma}S^{\alpha\beta}}{m^2 + \dfrac{1}{4}R_{\alpha\beta\gamma\delta}S^{\alpha\beta}S^{\gamma\delta}}   \right) \ , \label{MP-Velocity}
\end{align}
in which $ \bar{Z} \equiv P_{\lambda} U^{\lambda} $ is a constant scalar quantity.

The dynamics of an extended rotating object is governed by a coupled system of equations for which no closed-form analytical solution is known. To address this, several approximation schemes have been proposed, such as those introduced by Tod, de Felice, and Calvani \cite{Tod:1976}, Mashhoon and Singh \cite{Mashhoon:2006}, and a further approach by Singh \cite{Singh:2008}. In the first approach, authors analyzed the motion of spinning test bodies in the gravitational field of a rotating black hole, restricting their study to the pole--dipole approximation and to equatorial motion with the spin oriented perpendicular to the plane of motion. Mashhoon and Singh\cite{Mashhoon:2006}, developed a first order approximation method for the influence of spin on the motion of spinning test masses in a gravitational field. Finally, Singh\cite{Singh:2008} employs a power series expansion of the particle's spin to analytically describe the dynamics of a classical spinning particle in the context of Mathisson--Papapetrou--Dixon formalism. Another study of solving the equations of motion of an extended body at linear approximation for a monochrome gravitational can be found in the work of Mohseni and Sepanji\cite{Mohseni:2000}.  

In the following section, we seek for the solution of equations of motion of an extended system in a perturbed flat spacetime. The perturbation, is introduced via a gravitational wave pulse.

%%%%%%%%%%%%%%%%%%%%%%%%%%%%%%%%%%%% Section III
\section{Gravitational Wave Memory and the Extended Body Dynamics}
\label{Sec:Pert.}
In order to investigate the dynamics of an extended body in the gravitational field, we develop a perturbative scheme here. As the gravitational field we choose a gravitational wave pulse, as a wave packet is more realistic than a monochrome wave as studied partly for another purpose in \cite{Mohseni:2000}. 
 
Gravitational waves can be treated as a small perturbation ($h_{\mu\nu}$) to the flat spacetime, represented by the Minkowski metric $\eta_{\mu\nu} = (+1,-1,-1,-1)$.  The perturbed spacetime metric is then $g_{\mu\nu} = \eta_{\mu\nu} + h_{\mu\nu} $.  By applying appropriate gauge conditions in the linearized theory, redundant components can be removed, leaving only the two physical degrees of freedom. These correspond to the two standard polarization modes of the gravitational waves, i.e. the plus ($+$) and cross ($\times$) polarizations, which are typically the most useful basis for analyzing gravitational waves. 

In the Cartesian coordinates, 
\begin{equation}
g_{\mu\nu}^{\text{(Cartesian)}}=
\begin{pmatrix}
1 & 0 & 0 & 0 \\
0 & -1- \epsilon h_{+}(x^{\mu})  & \epsilon h_{\times}(x^{\mu})  & 0 \\
0 & \epsilon h_{\times}(x^{\mu})  & -1 + \epsilon h_{+}(x^{\mu})  & 0 \\
0 & 0 & 0 & -1
\end{pmatrix}
\end{equation}
where the infinitesimal parameter $\epsilon \ll 1$ is introduced for tracking the perturbation order, and the functions $h_+ (x^\mu)$ and $h_\times (x^\mu)$ encode the two polarization modes of the wave. Then, the Christoffel symbols and the spacetime curvature can be expressed as
\begin{align}
& \Gamma^{\lambda}_{\mu\nu} = \dfrac{\epsilon}{2} \eta^{\lambda\rho} \left( \partial_{\mu} h_{\rho\nu} +\partial_{\nu} h_{\rho\mu} - \partial_{\rho} h_{\mu\nu}  \right) \equiv \epsilon \tilde{\Gamma}^{\lambda}_{\mu\nu} \ , \label{Per-Connection}\\
&R_{\alpha\beta\mu\nu} = \dfrac{\epsilon}{2} \left( \partial_{\beta}\partial_{\mu}h_{\alpha\nu} +  \partial_{\alpha}\partial_{\nu}h_{\beta\mu} - \partial_{\alpha}\partial_{\mu}h_{\beta\nu} -  \partial_{\beta}\partial_{\nu}h_{\alpha\mu} \right) \equiv \epsilon \tilde{R}_{\alpha\beta\mu\nu}  \label{Per-Riemann}
\end{align}
up to the first order.

To examine the extended body dynamics, Mathisson--Papapetrou equations ((\ref{MP-Momentum})--(\ref{MP-Velocity})) can also be expanded up to the first order in $\epsilon$. To achieve this, the first and the second moments of the energy--momentum tensor, given by equations \eqref{1st-Moment} and \eqref{2nd-Moment}, and its four-velocity can be consistently expanded in powers of $\epsilon$ as
\begin{align}
& P^{\mu} = P^{\mu}_{(0)} + \epsilon P^{\mu}_{(1)} \ , \\
& S^{\mu\nu} = S^{\mu\nu}_{(0)} + \epsilon S^{\mu\nu}_{(1)} \ , \\
& U^{\mu} = U^{\mu}_{(0)} + \epsilon U^{\mu}_{(1)} 
\end{align}
where $P^{\mu}_{(1)}$, $S^{\mu\nu}_{(1)}$, and $ U^{\mu}_{(1)} $ denote the first order contributions to the momentum, spin, and velocity, respectively. These three, besides their nonperturbed counterparts $P^{\mu}_{(0)} $,  $S^{\mu\nu}_{(0)} $, and $U^{\mu}_{(0)}$ should satisfy equations \eqref{MP-Momentum} to \eqref{MP-Velocity}. It is straightforward to show that the resulting equations are given by
\begin{align}
& \dot{P}_{(0)\nu} = 0 \ , \label{MomentumD0} \\
& \dot{P}_{(1)\nu} = \dfrac{\bar{Z}}{m^2} \left(P^{\mu}_{(0)} P_{(0)\alpha} \tilde{\Gamma}^{\alpha}_{\nu\mu}  - \dfrac{1}{2} S_{(0)}^{\alpha\mu} P_{(0)}^{\beta} \tilde{R}_{\alpha\mu\nu\beta} \right) \ , \label{MomentumD1}
\end{align}
and
\begin{align}
& \dot{S}_{(0)\alpha\mu} = 0 \ , \label{SpinD0} \\
& \dot{S}_{(1) \alpha\mu} = \dfrac{\bar{Z}}{m^2} \left\{ P^{\nu}_{(0)} \left( S_{(0) \beta\mu} \tilde{\Gamma}^{\beta}_{\alpha\nu} +  S_{(0) \alpha\beta} \tilde{\Gamma}^{\beta}_{\nu\mu}  \right)  +  \dfrac{1}{m^2} S_{(0)}^{\lambda\delta} P^{\nu}_{(0)} \tilde{R}^{~~~\beta}_{\lambda\delta ~ \nu} \left( S_{(0) \alpha\beta}  P_{(0) \ \mu} + S_{(0) \beta\mu}  P_{(0) \alpha} \right) \right\} \ ,  \label{pinSD1}
\end{align}
and
\begin{align}
& U^{\mu}_{(0)} = \dfrac{\bar{Z}}{m^2} P_{(0)}^{\mu}  \ , \label{Velocity0} \\
& U^{\mu}_{(1)} =  \dfrac{\bar{Z}}{m^2} \epsilon \left(P_{(1)}^{\mu}  + \dfrac{1}{2m^2}  S_{(0)}^{\mu\nu} P^{\sigma}_{(0)} S_{(0)}^{\alpha\beta}\tilde{R}_{\nu\sigma\alpha\beta}  \right) \ . \label{Velocity1}
\end{align}
Here, a dot over any quantity represents differentiation with respect to the proper time $\tau$. Note that, since the unperturbed background is flat, any curvature term appears at the first order, and is denoted with a tilde to avoid confusion.

Since the unperturbed values of $P^{\mu}_{(0)} $,  $S^{\mu\nu}_{(0)} $, and $U^{\mu}_{(0)}$ act as nonhomogeneous sources in the dynamical equations for $P^{\mu}_{(1)}$, $S^{\mu\nu}_{(1)}$, and $ U^{\mu}_{(1)} $, we begin with solving the equations \eqref{MomentumD0}, \eqref{SpinD0}, and \eqref{Velocity0}. 
The first two equations impose constraints on $P^{\mu}_{(0)} $ and  $S^{\mu\nu}_{(0)} $ ensuring that they remain fixed at their initial values. To simplify things we choose a coordinate system in which
\begin{equation}
P_{(0)\mu} = (1,0,0,0) \ ,
\label{P-IntVa}
\end{equation}
and thus
\begin{equation}
U_{(0)\mu} =  \frac{\bar{Z}}{m^2} (1,0,0,0) \ .
\label{P-IntVa}
\end{equation}
To fix the spin orientation, we consider a rotating object (at zeroth order) confined to the $XZ$-plane, without any loss of the generality. Then,
\begin{equation}
S_{(0)\mu\nu} =
\begin{pmatrix}
0 & 0 & 0 & 0 \\
0 & 0 & S_z  & 0 \\
0 & -S_z & 0 & S_x \\
0 & 0  & -S_x & 0
\end{pmatrix} \ ,
\label{S0}
\end{equation}
where 
\begin{equation}
S_x=s \sin \alpha, \quad S_z=s \cos \alpha \ ,
\end{equation}
with $s$ denoting the spin length. 
These initial conditions ensure that the orthogonality condition  $S_{(0)\mu\nu} P_{(0)}^{\nu} = 0$ is fulfilled. 
 
Substituting these three into equations  \eqref{MomentumD1}, \eqref{pinSD1}, and \eqref{Velocity1}, while taking into account the equations \eqref{Per-Connection} and \eqref{Per-Riemann}  which represent the contribution of the gravitational wave along the $z$-axis to the object dynamics. This leads to the following relations
\begin{align}
& \dot{P}_{(1)\mu} = \left( 0, \epsilon f_1(t,z), \epsilon f_2(t,z), 0 \right) \ , \\
& \dot{S}_{(1)\mu\nu} =
\begin{pmatrix}
0 & \epsilon g_{01} (t,z) & \epsilon g_{02} (t,z) & \epsilon g_{03} (t,z) \\
-\epsilon g_{01}(t,z) & 0 & 0  & \epsilon g_1(t,z) \\
\epsilon g_{02}(t,z) & 0 & 0 & \epsilon g_2(t,z) \\
\epsilon g_{03}(t,z) & -\epsilon g_1(t,z)  & -\epsilon g_2(t,z) & 0
\end{pmatrix} \ , \\
& U^{\mu}_{(1)} = \left(0, \epsilon( P^1_{(1)} + e_1(t,z) ), \epsilon ( P^2_{(1)}+  e_2(t,z)),  \epsilon e_3(t,z) \right)
\end{align}
where for the last one we used the relation $\dv*{t}{\tau} = 1 $, since $U^{0} = 1$.

The form of the functions $f_1(t,z)$, $f_2(t,z)$, $g_1(t,z)$, $g_2(t,z)$, $g_{0i}(t,z)$, and $e_i(t,z)$ for $i=\{1,2,3\}$, depend on the gravitational wave fields $h_+ (t,z)$ and $h_{\times} (t,z)$.
For a Gaussian wave pulse  with amplitudes $H_+$ and $H_{\times}$,
\begin{align}
	& h_+ (t,z) = H_+ \  e^{-\frac{(z-c t)^2}{2\sigma^2}} e^{ik(z - ct)} \ , \\
	& h_{\times} (t,z) = H_{\times} \ e^{-\frac{(z-c t)^2}{2\sigma^2}} e^{ik(z - c t)} \ ,
\end{align}
these functions take the forms
\begin{eqnarray}
&	\{ f_1,f_2 \} =\{ \frac{S_x H_{\times}}{2 \sigma^4}  \mathcal{F}(v), \frac{S_x H_{+}}{2 \sigma^4} \mathcal{F}(v) \} \ , \\
&	\{ g_{01} ,g_{02} ,g_{03}  \} =\{ - \frac{S_x S_z H_+}{\sigma^4} \mathcal{F}(v), \frac{S_x S_z H_{\times}}{\sigma^4}  \mathcal{F}(v), \frac{S_x^2 H_+}{\sigma^4} \mathcal{F}(v) \}  \ , \\
&	\{ g_1 ,g_2  \} =\{ \frac{S_x H_{\times}}{2 \sigma^2}  \mathcal{G}(v), \frac{S_x H_+}{2 \sigma^2}  \mathcal{G}(v) \} \ , \\
&	\{ e_1 ,e_2 ,e_3  \} =\{ -\frac{S_x S_z H_+}{2 \sigma^4}  \mathcal{F}(v), \frac{S_x S_z H_{\times}}{2 \sigma^4}  \mathcal{F}(v),  \frac{S_x^2 H_+}{2 \sigma^4}  \mathcal{F}(v) \}  \ ,
\end{eqnarray}
where  $v \equiv t-z$  and
\begin{align}
	&  \mathcal{F}(v) \equiv e^{-v^2/2 \sigma^2} \big (  (\sigma^4+\sigma^2-v^2)\cos v - 2\sigma^2 v \sin v \big) \ , \\
	&  \mathcal{G}(v) \equiv e^{-v^2/2 \sigma^2} \left( \sigma^2 \sin v + v\cos v  \right) \ .
\end{align}

The complete set of equations needed to solve for obtaining the dynamics of an extended body as a gravitational wave pulse passes is thus
\begin{align}
	& \dv{P_{(1)1}}{t} = f_1 (t,z) , \quad\quad\quad\quad\quad  \dv{P_{(1)2}}{t} = f_2 (t,z) \\
	& \dv{S_{(1)x}}{t} = g_2 (t,z) , \quad\quad\quad\quad\quad \dv{S_{(1)y}}{t} = g_1 (t,z) \\
	& \dv{X}{t} = \epsilon P_{(1)}^1 + \epsilon e_1 (t,z)  , \quad\quad\   \dv{Y}{t} = \epsilon P_{(1)}^2 + \epsilon e_2 (t,z)  , \quad\quad\ \dv{Z}{t} = \epsilon e_3 (t,z)\ . 
\end{align}

The above equations can be solved numerically. In what follows the results for two cases of the wave being essentially $+$-type ($\varsigma \equiv H_+/H_{\times}\gg 1$), or being essentially $\times$-type ($\varsigma\ll 1$), are discussed. The values of normalized pulse width $k\sigma=10$, initial spin orientation $\alpha=35^{\circ}$, and $\epsilon =0.1$ are chosen in the plots and we set $c=1$ and everything is normalized by $k$, the wave number.

\begin{itemize}
\item For a wave packet being essentially $+$-type we choose $\varsigma=10$. The spin orientation behaves as is shown in figures (\ref{theta-HpTHc-10}) and (\ref{phi-HpTHc-10}).

As it can be seen from the plots the angle between the spin and the direction of the propagation of the wave, comes back to its original value after several oscillations. But the spin orientation in the plane orthogonal to the wave direction experiences a small change after oscillations. It means that after some nutation and precession there would be some memory in the $\phi$-direction.

The motion of the specific point of the body is also shown in figures (\ref{TraMo-HpTHc-10}) and (\ref{LonMo-HpTHc-10}). One can see again a jump in the $z$-component and an spiral motion in the transverse plane. Therefore, there is gravitational hysteresis both in the linear and spin motion of the body.

\begin{figure}[H]
\centering
\includegraphics[width=0.9\textwidth]{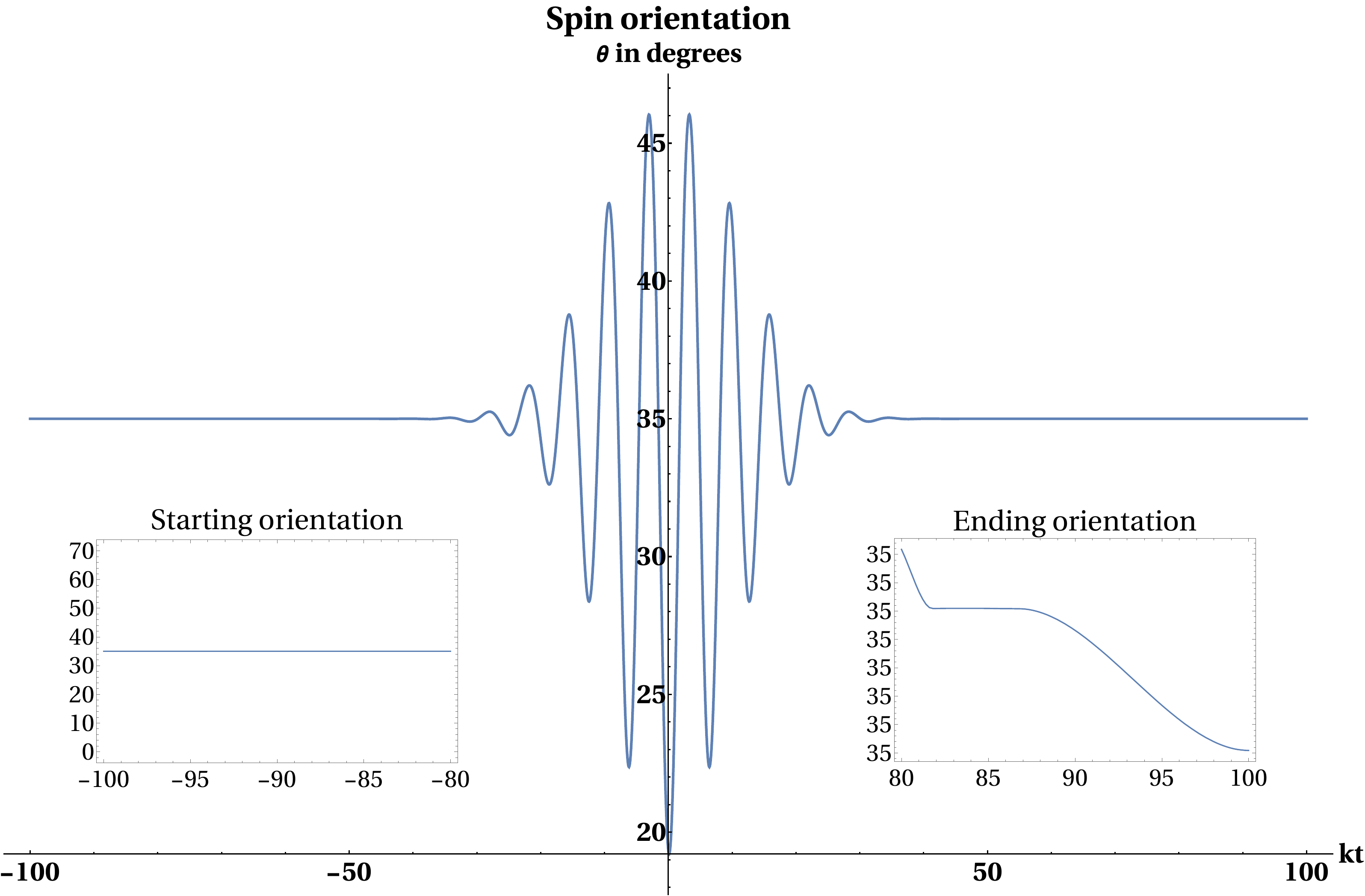}
\caption{Change in $\theta$ orientation for $\varsigma=10$.}
\label{theta-HpTHc-10}
\end{figure}

\begin{figure}[H]
\centering
\includegraphics[width=0.9\textwidth]{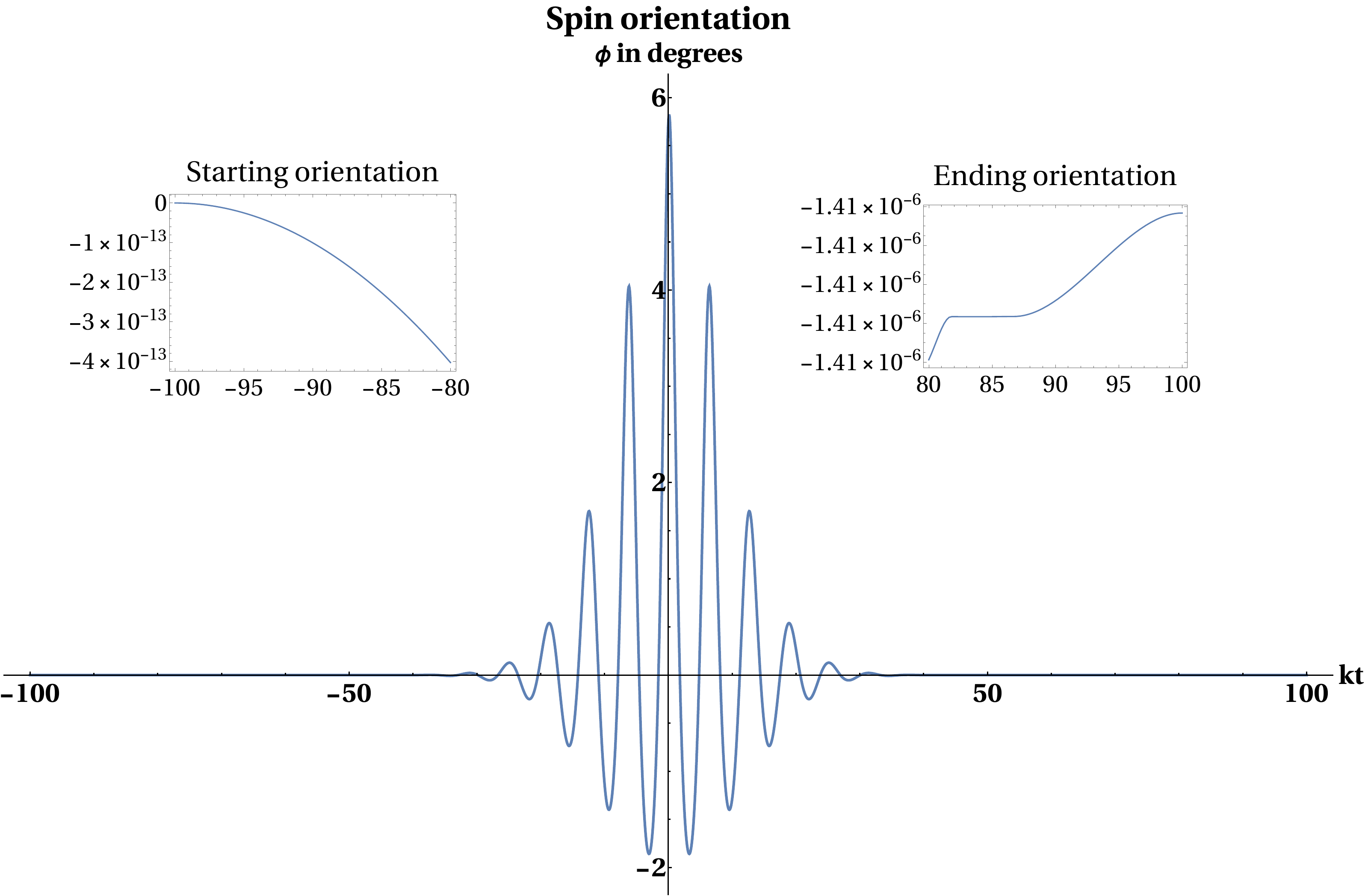}
\caption{Change in $\phi$ orientation for $\varsigma=10$.}
\label{phi-HpTHc-10}
\end{figure}

\begin{figure}[H]
\centering
\includegraphics[width=0.5\textwidth]{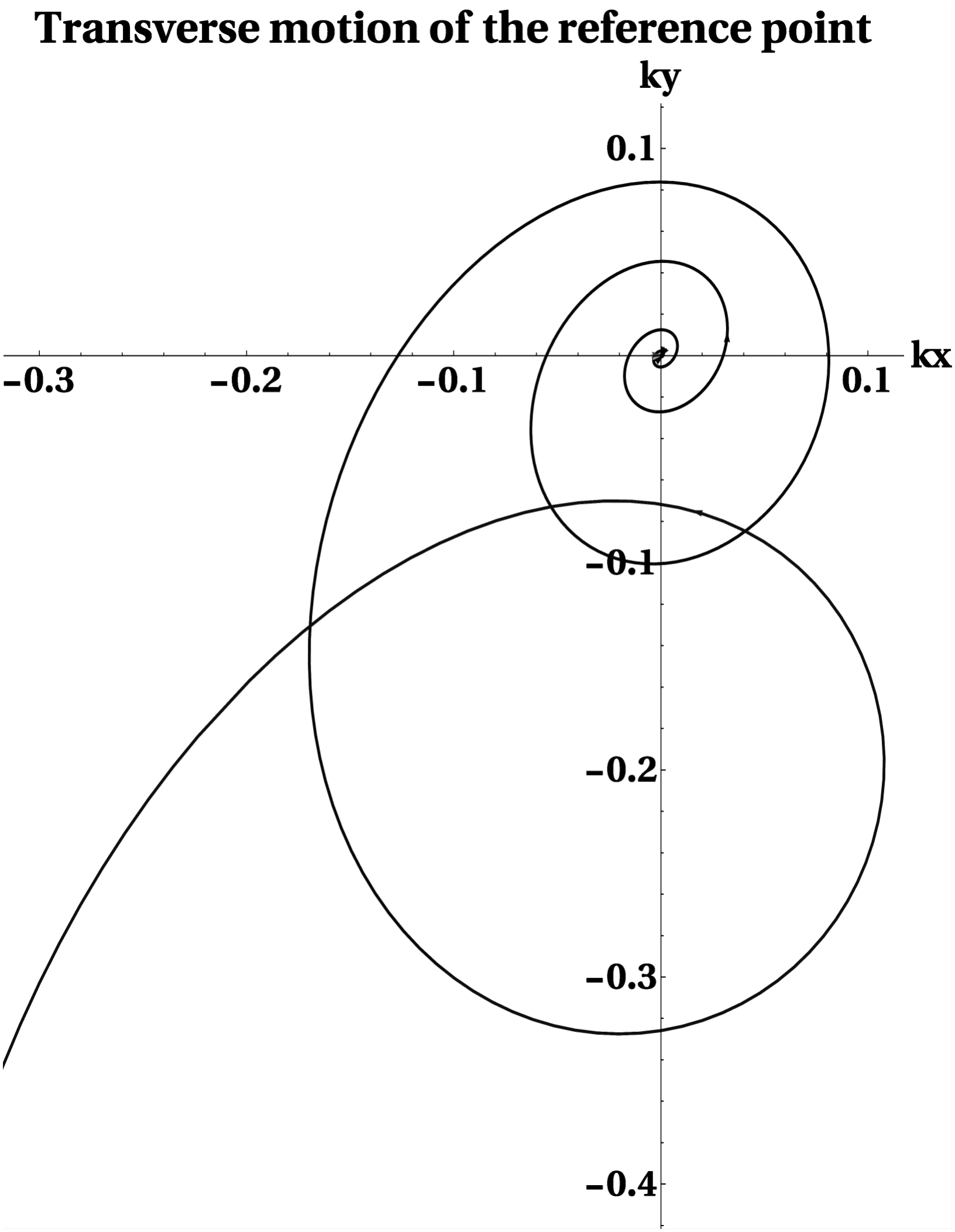}
\caption{Transverse motion of the reference point for $\varsigma=10$.}
\label{TraMo-HpTHc-10}
\end{figure}

\begin{figure}[H]
\centering
\includegraphics[width=0.9\textwidth]{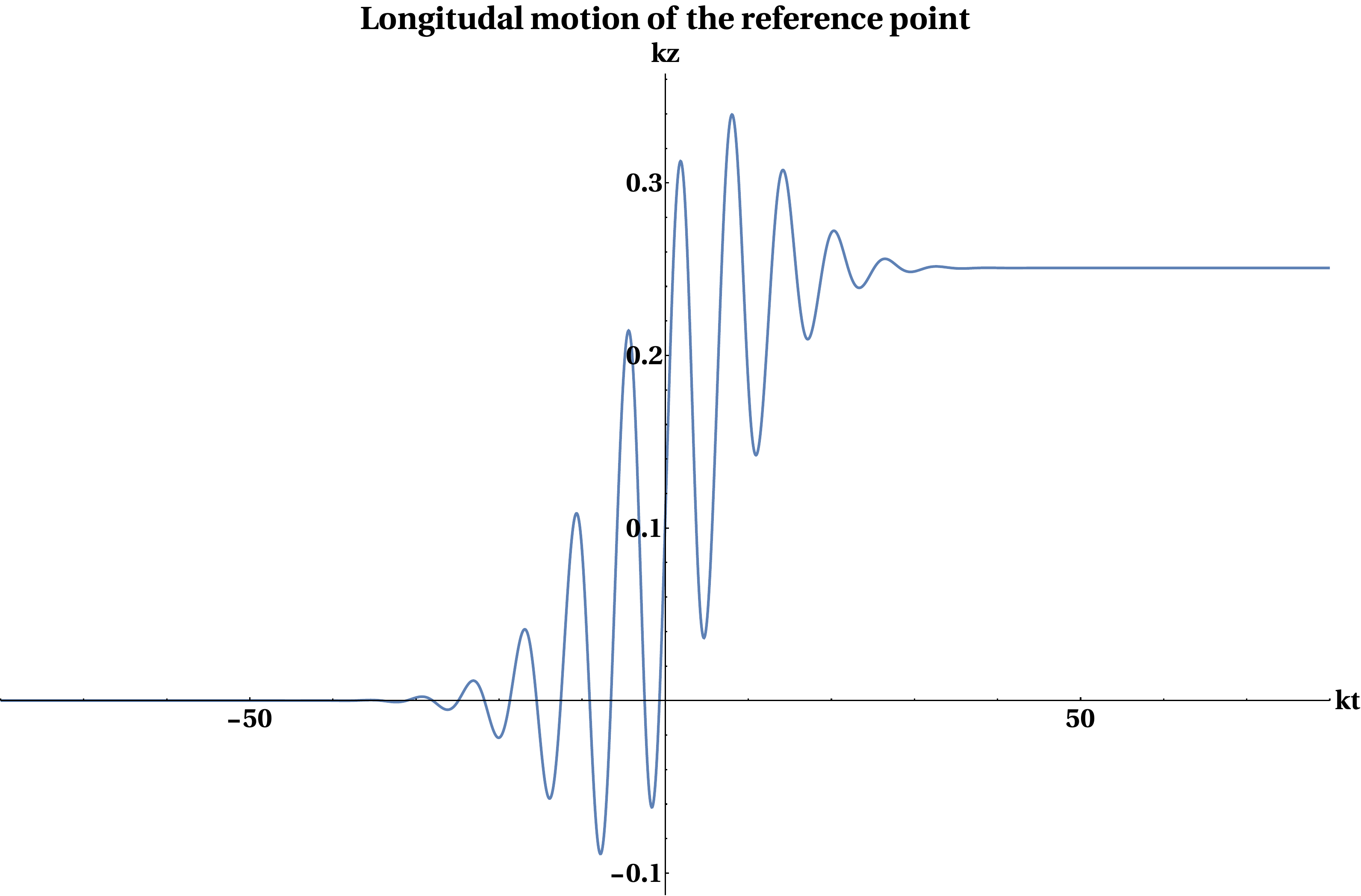}
\caption{Longitudal motion of the reference point for $\varsigma=10$.}
\label{LonMo-HpTHc-10}
\end{figure}

\item As the second case, we consider a wave packet that is essentially $\times$-type with $\varsigma=0.3$. The spin orientation is illustrated in figures (\ref{theta-HpTHc-0.3}) and (\ref{phi-HpTHc-0.3}).

Again we can observe that after some oscillations, the angle between the spin and the direction of the propagation of the wave, comes back to its original value while the spin orientation in the plane orthogonal to the wave direction shows memory.

Also the motion of the specific point of the body (see figures (\ref{TraMo-HpTHc-0.3}) and (\ref{LonMo-HpTHc-0.3})) has a similar behavior to the previous case. There is a smaller jump in the $z$-component and an spiral motion in the transverse plane. 

\begin{figure}[H]
\centering
\includegraphics[width=0.9\textwidth]{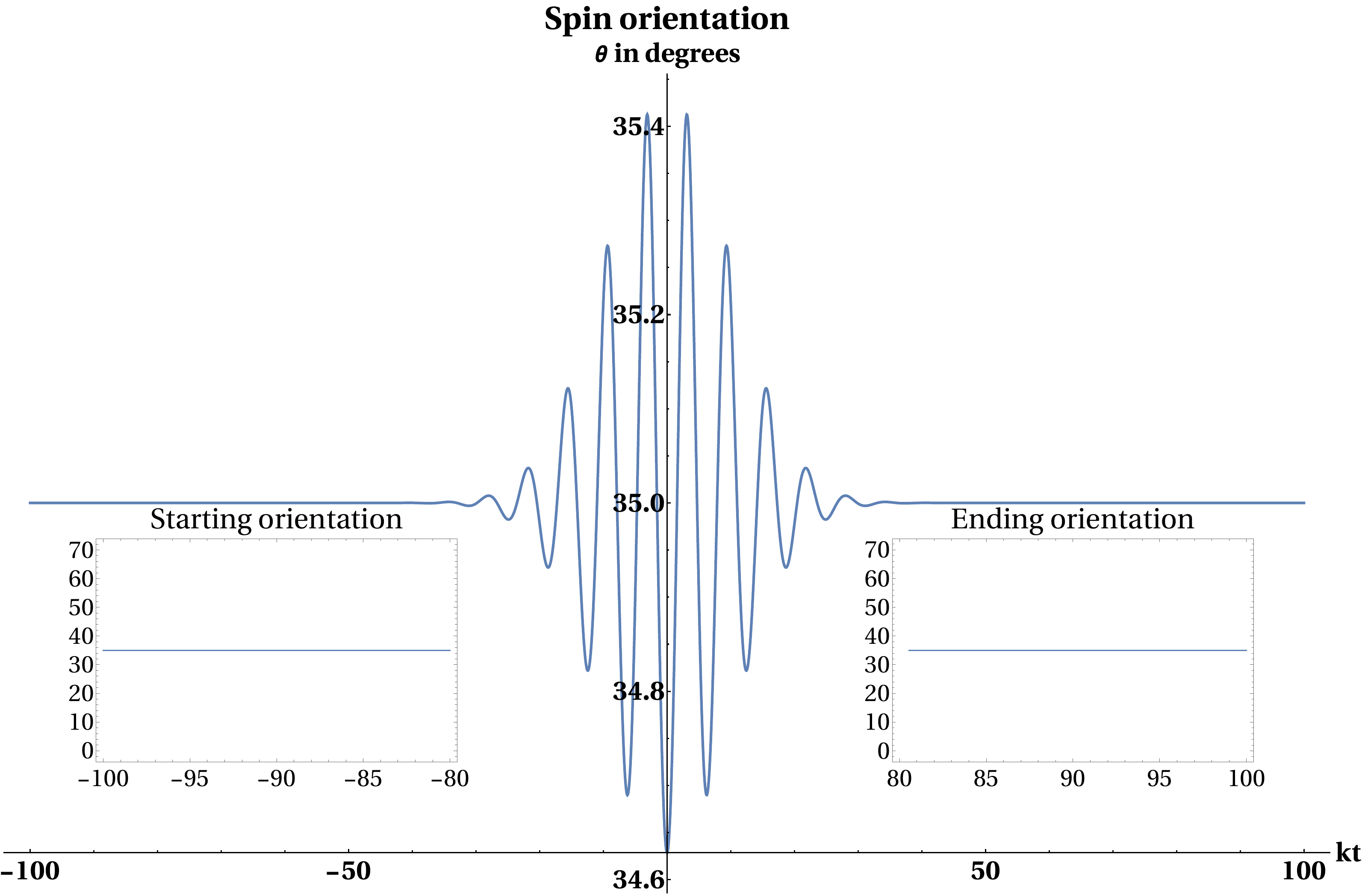}
\caption{Change in $\theta$ orientation for $\varsigma=0.3$.}
\label{theta-HpTHc-0.3}
\end{figure}

\begin{figure}[H]
\centering
\includegraphics[width=0.9\textwidth]{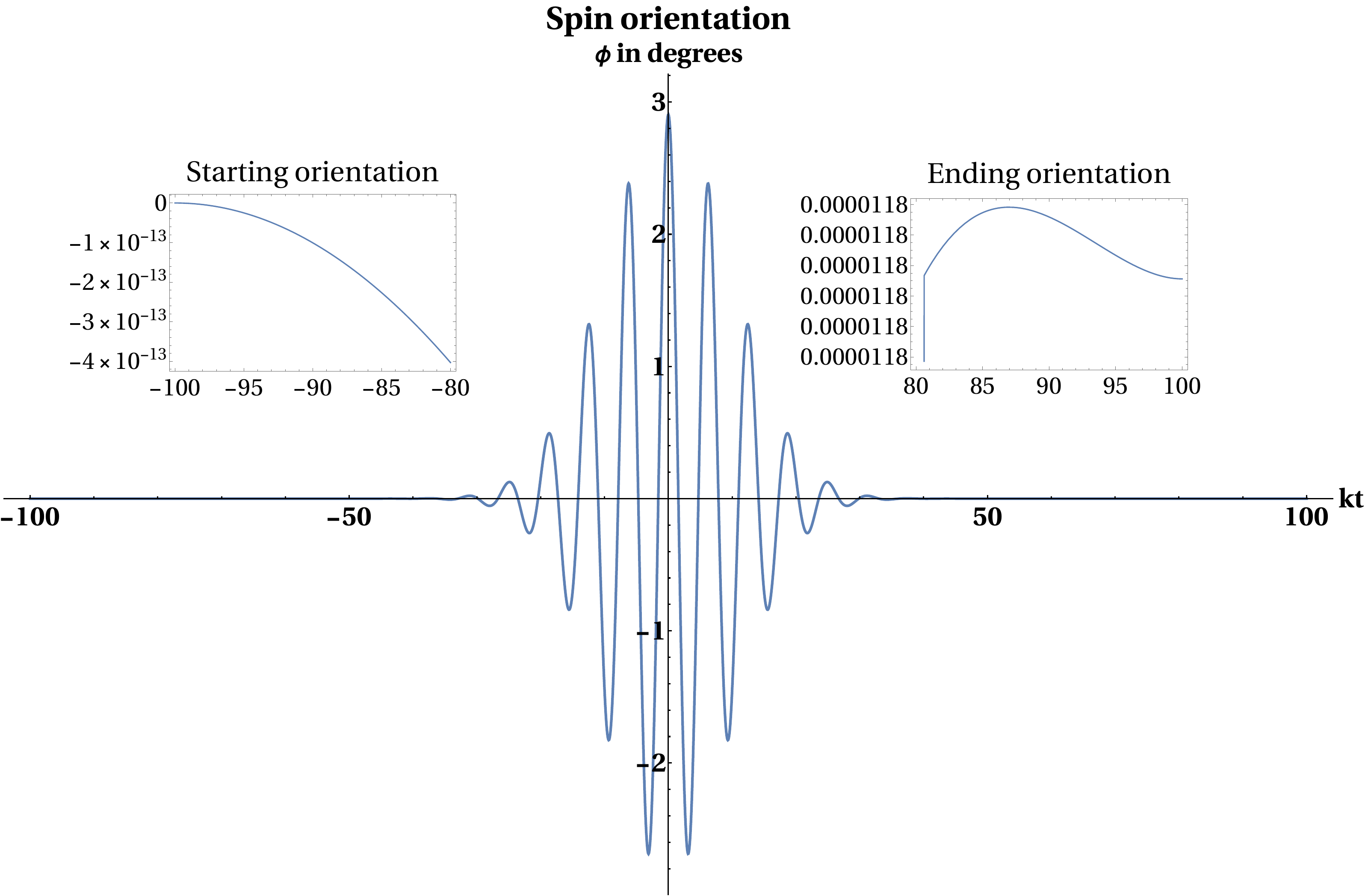}
\caption{Change in $\phi$ orientation for $\varsigma=0.3$.}
\label{phi-HpTHc-0.3}
\end{figure}

\begin{figure}[H]
\centering
\includegraphics[width=0.9\textwidth]{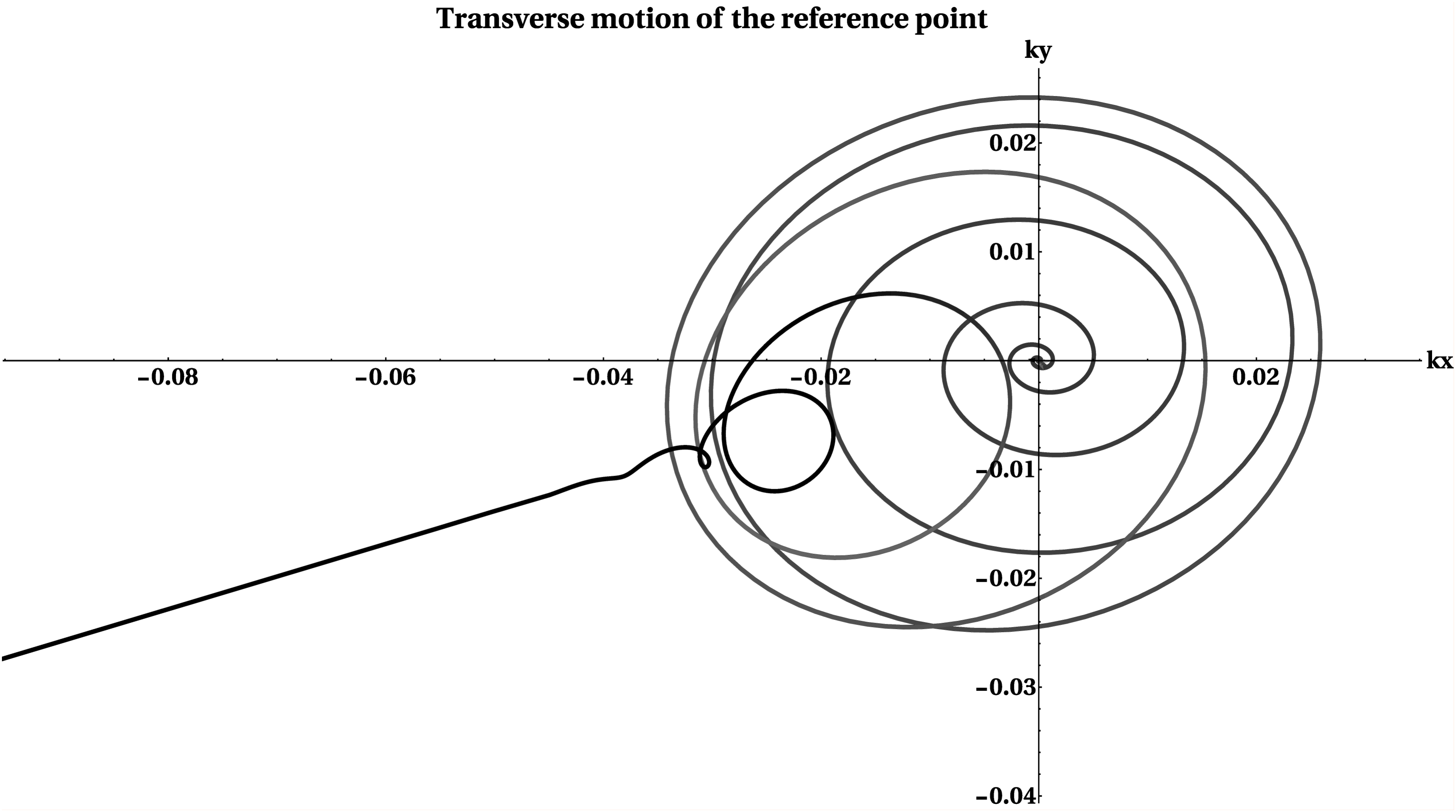}
\caption{Transverse motion of the reference point for $\varsigma=0.3$.}
\label{TraMo-HpTHc-0.3}
\end{figure}

\begin{figure}[H]
\centering
\includegraphics[width=0.9\textwidth]{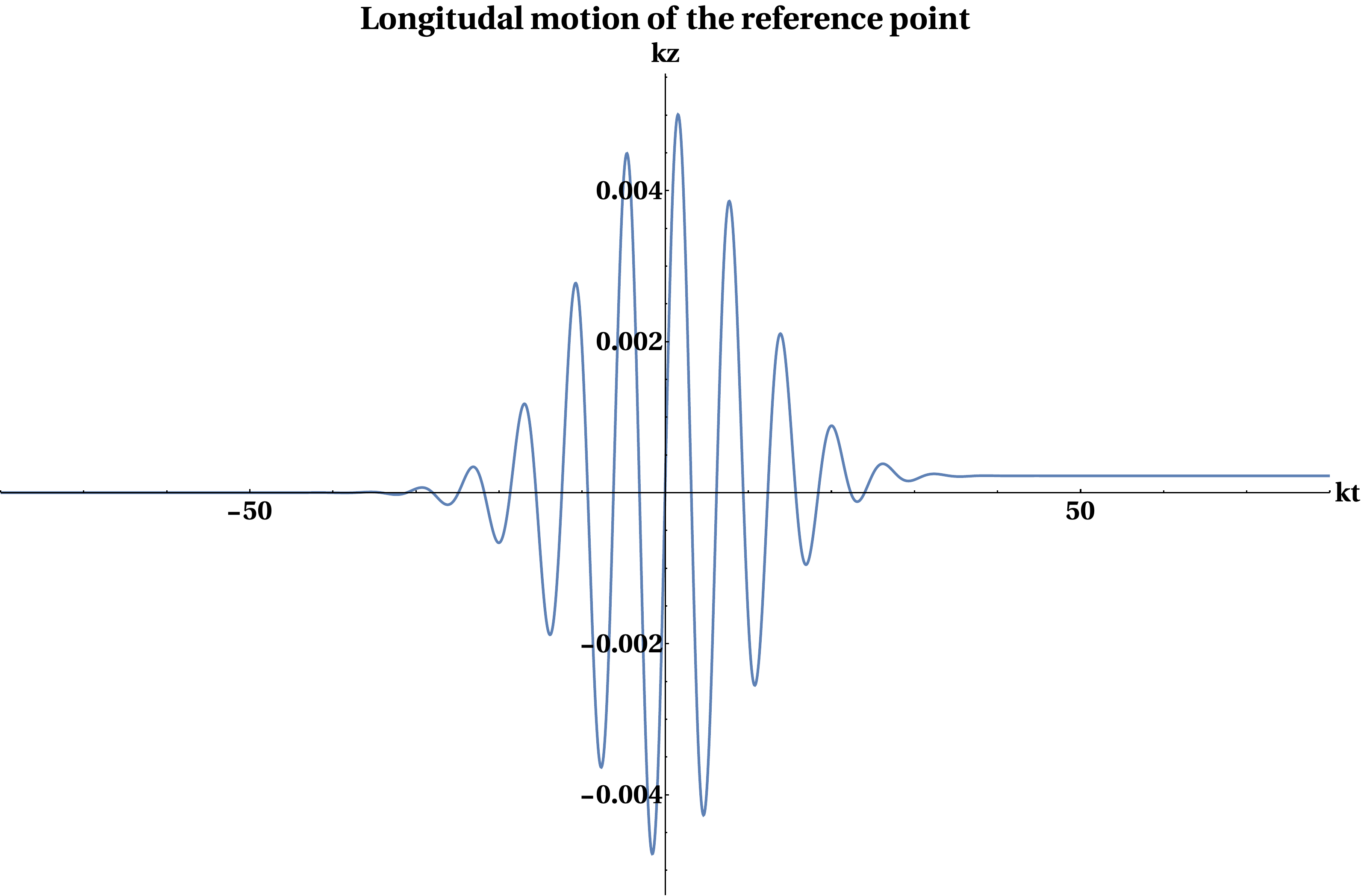}
\caption{Longitudal motion of the reference point for $\varsigma=0.3$.}
\label{LonMo-HpTHc-0.3}
\end{figure}

\end{itemize}

It can be concluded that the dynamics of an extended body memorizes the footprint of the passage of a gravitational wave pulse. This memory is present both in its linear and rotational motion.

\begin{figure}[H]
\centering
\includegraphics[width=0.9\textwidth]{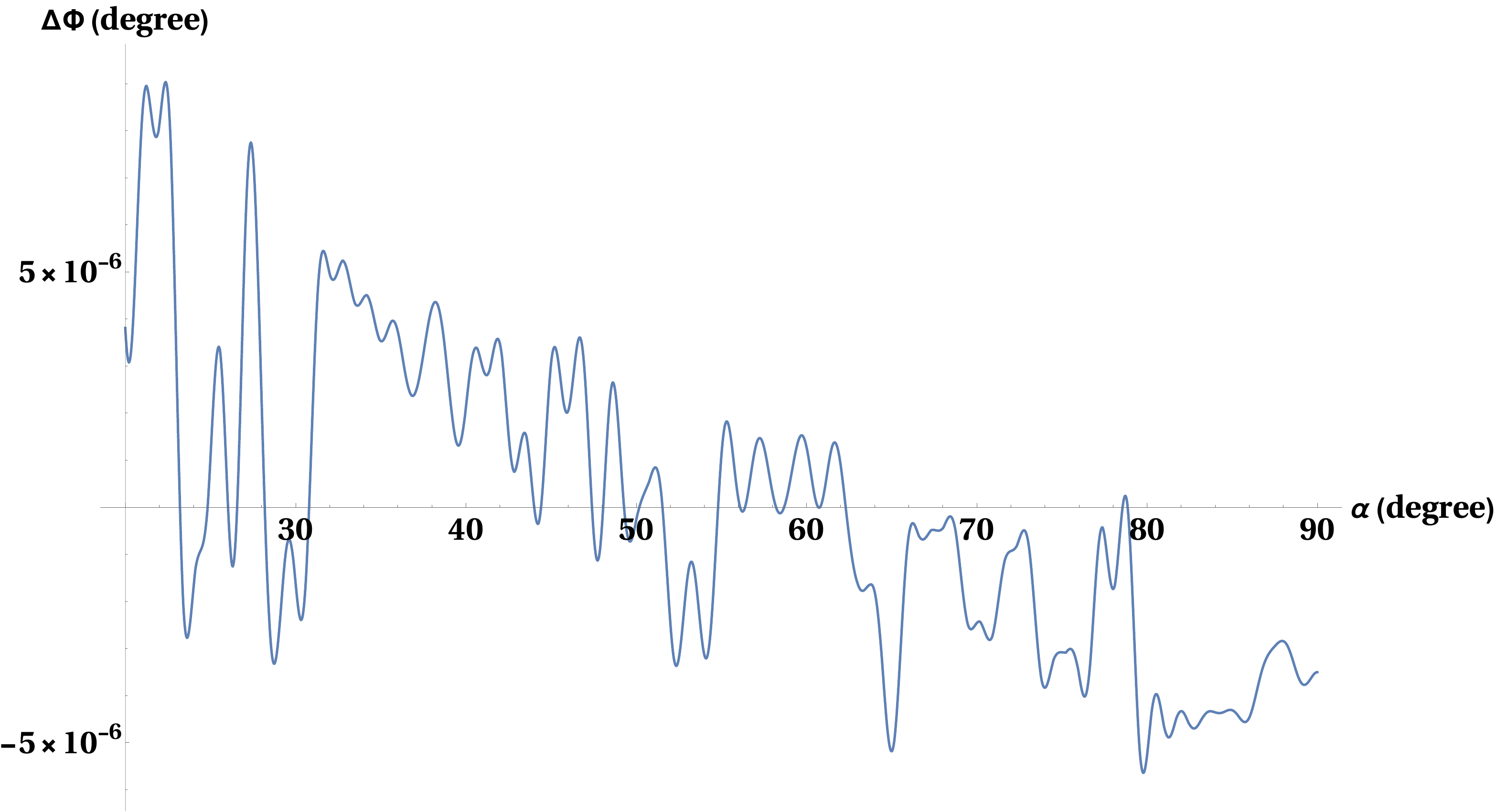}
\caption{Memory of $\phi$-angle of spin for $\varsigma=3$.}
\label{DeltaPhi}
\end{figure}

The above plots show the memory effect for a specific initial angle between spin and wave propagation direction. As the change in $\phi$-angle is a direct measure of memory, it is fruitful to investigate its dependence on the initial angle ($\alpha$) between spin and wave number. This is shown in figure (\ref{DeltaPhi}), for $\varsigma=3$ and $\epsilon =0.1$. It can be seen that $\Delta\phi$ has a highly oscillating nature. This is a general behavior for any choice of $\varsigma$.

In section \ref{Sec:Ensemble}, we will use an ensemble of extended bodies and study their thermodynamics. For such randomly oriented bodies, this highly oscillatory nature is integrated out leading to a smooth contribution to the thermodynamic quantities.

%%%%%%%%%%%%%%%%%%%%%%%%%%%%%%%%%%%% Section IV
\section{Non-linear Memory of the Gravitational Wave}
\label{Sec:NonlinearMemory}
As the gravitational wave pulse causes the body to oscillate both in its linear and spin motion, the object would radiate and contribute an additional radiation ($h_{ij}^{\text{rad}}$) to the total gravitational field. Therefore the wave itself, also experiences a memory effect. This can be called non-linear memory effect, that is, the memory sector of radiation generated by an extended rotating source in curved spacetime.

In particular, the second time derivative of the source quadrupole moment is directly related to the transverse--traceless (\textsc{tt}) part of the radiated field, falling off with distance as  $1/r$ \cite{Maggiore:2007ulw}, 
\begin{equation}
[ h_{ij}^{\text{rad}}]^{\textsc{tt}} = \dfrac{1}{r} \dfrac{2G}{c^4} \Lambda_{ij,kl} (\hat{\vb{n}})  \ddot{Q}_{ij} (t-\dfrac{r}{c})
\end{equation}
where the transverse--tracelss operator is
\begin{equation}
\Lambda_{ij,kl} (\hat{\vb{n}}) = \delta_{ik} \delta_{jl} - \dfrac{1}{2} \delta_{ij} \delta_{kl} -n_j n_l \delta_{ik} -n_i n_k \delta_{jl} + \dfrac{1}{2} n_k n_l \delta_{ij} + \dfrac{1}{2} n_i n_j \delta_{kl} + \dfrac{1}{2} n_i n_j n_k n_l  \ ,
\end{equation}
and for an object with the mass moment $M_{ij}$,
\begin{align}
Q^{ij} & \equiv M^{ij} - \dfrac{1}{3} \delta^{ij} M_{kk}  \nonumber \\
& = \int \dd[3]{x} \rho(t,\vb{x}) (x^i x^j - \dfrac{1}{3} r^2 \delta^{ij}) \ .
\end{align}
These considerations lead to the well known relations in the spherical coordinates \cite{Maggiore:2007ulw}
\begin{align}
 h_+^{\text{rad}}(t,r,\theta,\phi) = &  \ddot{M}_{11} \left(\cos^2\phi-\sin^2\phi \cos^2\theta\right)+ \ddot{M}_{22} \left(\sin^2\phi-\cos^2\phi\cos^2\theta\right) \nonumber \\ 
 & -\ddot{M}_{33}\sin^2\theta -\ddot{M}_{12}\sin2\phi \left(1+\cos^2\theta\right) + \ddot{M}_{13}\sin\phi\sin2\theta + \ddot{M}_{23} \cos\phi\sin2\theta \ ,   \\
h_\times^{\text{rad}} (t,r,\theta,\phi) = & \left(  \ddot{M}_{11} -\ddot{M}_{22} \right) \sin2\phi\cos\theta + 2\ddot{M}_{12}\cos2\phi \cos\theta -2 \ddot{M}_{13}\cos\phi\sin\theta \nonumber \\ 
 &  + 2 \ddot{M}_{23} \sin\phi\sin\theta \ .
\end{align}
To have an order of magnitude of the radiated field, one can observe that there is a maximum direction for which
\begin{equation}
h_+^{\text{rad}} \ , h_{\times}^{\text{rad}} \sim \dfrac{1}{r} \dfrac{G}{c^4} \ddot{M}_{ij}\ .
\end{equation}

For an extended body of mass $m$, the mass moment is $M_{ij} \sim m x^2$, where $x$ characterizes the displacement. Its second time derivative is determined by the dynamical trajectory of $x$ obtained in the previous section. For linear motion there would be two terms $M_{ij} \sim mv^2$ and $M_{ij} \sim m l  a^2$, where $v$ and $a$ denoting the linear velocity and acceleration, respectively, and $l$ is the characteristic size of the object. On the other hand, for rotational motion, $M_{ij} \sim m l^2 \omega^2$, where the angular velocity $\omega$ may correspond to $\omega_{\theta}$ or $\omega_{\phi}$. Thus, in general the radiated field would consist of four terms
\begin{equation}
h_{+/\times}^{\text{rad}} = h^{(\text{v})} +  h^{(\text{a})} + h^{(\theta)} + h^{(\phi)}
\label{h-terms}
\end{equation}
where 
\begin{equation}
\{ h^{(\text{v})},  h^{(\text{a})}, h^{(\theta)}, h^{(\phi)} \} \sim \left(\dfrac{Gm}{c^4} \right)  \{ v^2,  la, \omega_{\theta}^2, \omega_{\phi}^2 \} \ .
\end{equation}

To see which term in equation (\ref{h-terms}) is dominant, we can estimate $v$, $a$, $\omega_\theta$, and $\omega_\phi$ as $\delta x k c/2\pi$, $\delta x k^2 c^2/4\pi^2$, $\delta\theta k c/2\pi$, and $\delta\phi k c/2\pi$, respectively. It is a very simple task to see from the numerical results that for large wavelengths ($l\ll\lambda$) $h^{(\text{v})}$ is the most important term, while for small wavelengths ($l\gg\lambda$) either $h^{(\theta)}$ or $h^{(\phi)}$ dominates.

Noting that up to the approximation we have adopted here the body motion is linearly dependent on the incoming wave amplitude, we can conclude that

\begin{equation}
h_{+/\times}^{\text{total}} (t,z)= h_{+/\times}^{\text{in}} (t,z) + h_{+/\times}^{\text{rad}} (t,z)
\end{equation}
with 

\begin{equation}
 h_{+/\times}^{\text{rad}} (t,z) = H_{+/\times}^{\text{rad}} \  e^{-\frac{(z-c t)^2}{2\sigma^2}} e^{ik(z - ct)} \ ,
\end{equation}
and that 
\begin{equation}
H^{\text{rad}}_{+/\times}  = H^{\text{in}}_{+/\times} \left(\dfrac{Gm}{4\pi^2 c^2 \varsigma^2 r} \right) 
  \begin{cases}
  1 & l \ll \lambda \ \\
  l^2 / \lambda^2  & l \gg \lambda \
\end{cases}
 \ .
\end{equation}

To figure out how much radiation the body adds to the incoming wave, consider the extrem case of an incoming gravitational wave produced by merging of two black holes, hitting the earth as the rotating extended body. Using the maximum frequency $\nu=10 ~\text{kHz}$ and mass and size of earth, we see that
\begin{equation}
\dfrac{H^{\text{rad}}_{+/\times}}{H^{\text{in}}_{+/\times}}  \sim \dfrac{2.7\times 10^{-3}}{r} \ .
\end{equation}
That is to say, in such a situation there would be a memory effect in the wave itself of order of about $\%0.27$ per unit distance from earth. Although this is very small, but it demonstrates the presence of non-linear memory of the incoming wave.

%%%%%%%%%%%%%%%%%%%%%%%%%%%%%%%%%%%% Section V
\section{Ensemble Analysis}
\label{Sec:Ensemble}
As a further step in our analysis of the dynamics of extended objects in the presence of a gravitational wave pulse, we explore the statistical and thermodynamic properties of an ensemble of them. Such a system may be realized by an ensemble of molecules considered as classical objects. 

\begin{figure}[htp]
\centering
\includegraphics[width=0.67\textwidth]{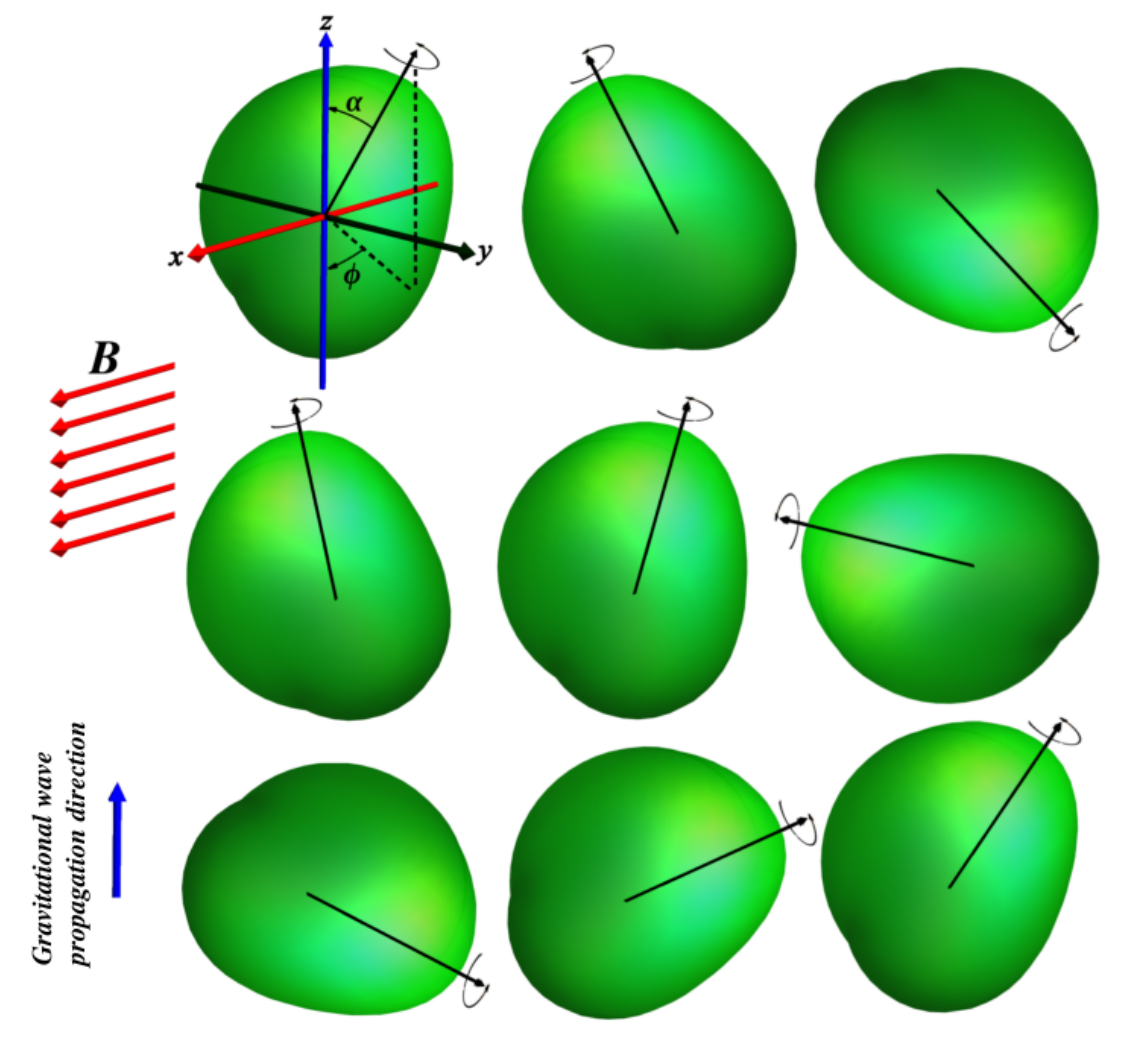} 
\caption{An ensemble of rotating objects with initial random orientations in constant magnetic field $\vb{B}=B\vu{x}$. A gravitational wave pulse run over the system in $z$-direction.}
\label{Ensemble}
\end{figure}

To begin, consider a collection of $N$ randomly oriented spinning objects, where the orientation of the $i$-th object is described by the angles $(\theta_i,\phi_i)$ for $i=1,\ldots, N$, as in figure (\ref{Ensemble}). There is a magnetic field in the $x$-direction, as shown in the figure, interacting with the magnetic moments of the objects. Denoting the system Hamiltonian as $H = H_0 + H_{\text{int}}(\theta_i,\phi_i)$ the partition function of the system is
\begin{equation}
\mathcal{Z}(T) \sim \int e^{-\beta H_{\text{int}}} \dd{\Omega}
\end{equation}
where $\beta \equiv  (k_B T)^{-1}$.
Following the passage of the gravitational wave pulse, each component of the system reorients to $(\theta_i+\Delta\theta_i, \phi_i+\Delta\phi_i)$. Consequently, the interaction Hamiltonian changes to $H_{\text{int}}(\theta_i+\Delta\theta_i, \phi_i+\Delta\phi_i)$. This shift from the initial configuration results in a change in the partition function, $\Delta\mathcal{Z}$, reflecting the memory of the ensemble about the wave.

Note that we implicitly assumed that the inter-molecular interaction balances the linear motion induced by the wave and only consider their rotation.

The values of $(\Delta\theta_i, \Delta\phi_i)$ depend on the system dynamics. In our model, the dynamics is governed by the Mathisson--Papapetrou equations. As stated in section \ref{Sec:Pert.}, numerical solutions reveal that $\Delta\theta_i = 0$.
Additionally, we assume that the wave propagates along the $z$-axis, which allows us to simplify the model by identifying the orientation with respect to the propagation direction  i.e. $\alpha$, with $\theta_i$. As a result, the effect of the initial orientations on the final distribution is encoded through the dependence of $\Delta\phi_i$ on $\theta_i$, as confirmed by the numerical results.

These considerations allow us to simplify the analysis by replacing the integration over $\theta_i$ with a discrete averaging. Therefore
\begin{equation}
 \mathcal{Z}(T) \sim \dfrac{\pi}{N} \sum_{i=1}^N \mathcal{Z}_i(T,\theta_i) \ ,
 \label{Z}
\end{equation}
where
\begin{equation}
\mathcal{Z}_i(T,\theta_i) = \int e^{-\beta H_{\text{int}}(\theta_i, \phi+\Delta\phi_i(\theta_i))} \sin{\theta_i} \dd{\phi} \ .
\label{Z-theta}
\end{equation}
To evaluate the partition function one assumes a random distribution of $\theta_i$ values. This enables us to evaluate $\Delta\mathcal{Z}$, change in the partition function due to the passage of the gravitational wave pulse. This represents the gravitational memory encoded into the partition function and thus in the thermodynamic quantities.

A realistic model for defining the Hamiltonian involves an ensemble of magnetic moments placed in a magnetic field. The interaction Hamiltonian is given by $ H_{\text{int}} \sim \mu \vb{B}\vdot \vb{S}$. Thus
\begin{equation}
H_{\text{int}} \sim \mu B S_{\perp} \cos{\phi}
\end{equation} 
where $S_{\perp} = \abs{\vb{S}} \sin{\theta_i} $. Using this Hamiltonian, we have 
\begin{equation}
\mathcal{Z}_i(T,\theta_i) = \int e^{- \frac{T_0}{T} \sin{\theta_i} \cos{(\phi+\Delta\phi_i(\theta_i))}} \sin\theta_i \dd{\phi}
\label{Z_i}
\end{equation}
where $T_0=\mu B \abs{\vb{S}}/k_B$ is a constant with the unit of temperature, the system characteristic temperature.  

\begin{figure}[H]
\centering
\includegraphics[width=0.75\textwidth]{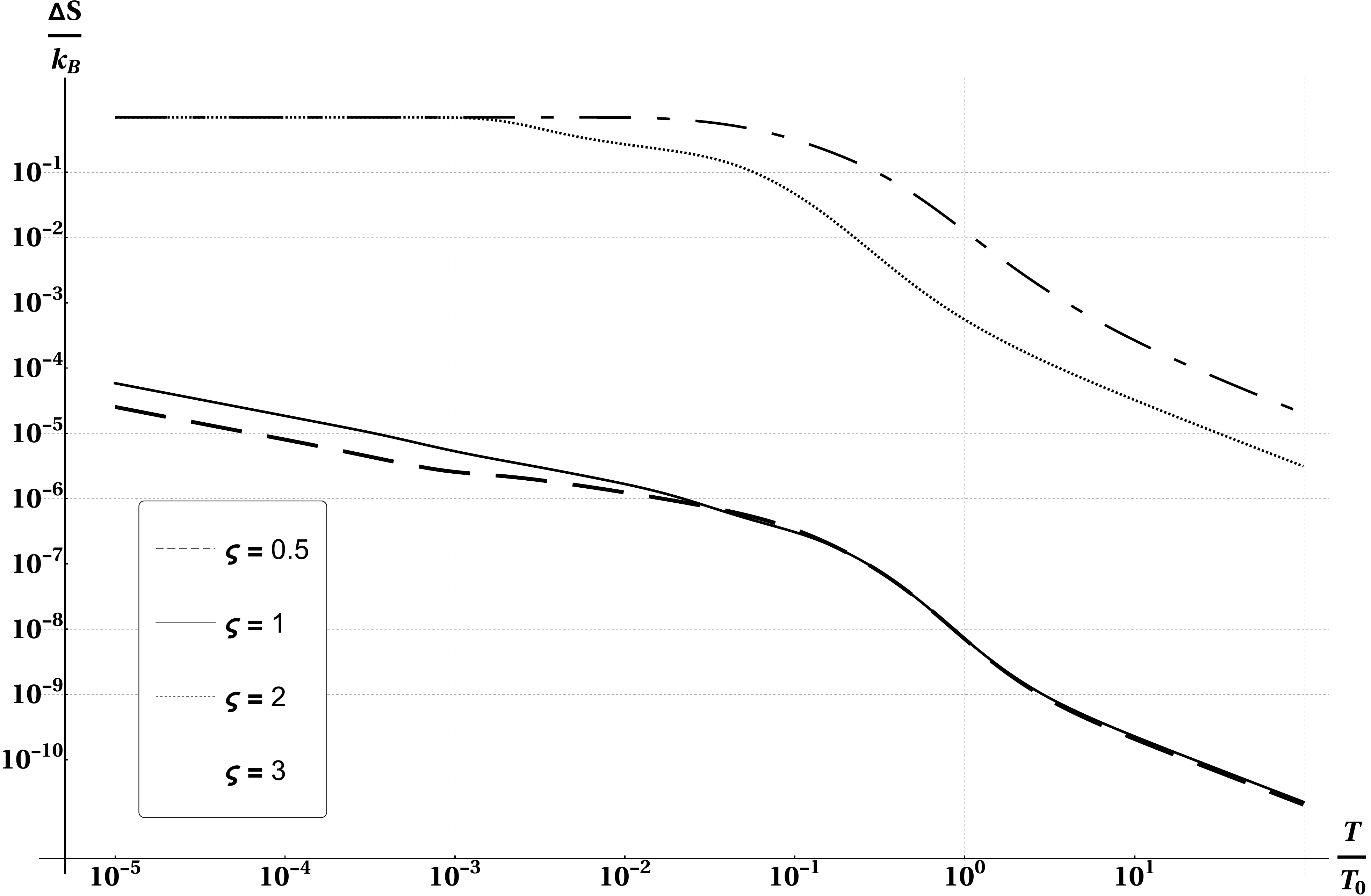} 
\caption{Footprint of the gravitational wave passage in the entropy.}
\label{Entropy}
\end{figure}

In order to obtain the change in the partition function, let's define $\zeta = \ln\frac{\cal{Z}}{\mathcal{Z}_0} $, where any subscript ``$0$" denotes the unperturbed value of the quantity, i.e. before the system interacts with the wave. Evaluating $\mathcal{Z}_i(T,\theta_i)$ using the numerical values of $\Delta\phi_i(\theta_i)$ from section \ref{Sec:Pert.} is straightforward. By averaging over all values of $\theta_i$ through equation \eqref{Z}, we can study the behavior of one of the most physically relevant quantities, namely the entropy ${\cal S}$. For  
\begin{equation}
\dfrac{{\cal S}}{k_B} = \beta U + \ln{\cal{Z}}
\end{equation}
where $U$ is the energy of the system, we have
\begin{equation}
\dfrac{\Delta {\cal S}}{k_B} = \beta \Delta U + \zeta \ ,
\end{equation}
with $\Delta {\cal S} \equiv {\cal S} - {\cal S}_0$ and $\Delta U \equiv U - U_0$. Since $\Delta U = \Delta\left ( \beta T \dv{\ln{\cal{Z}}}{T}\right )=\beta T \dv{\zeta}{T}$, we get
\begin{equation}
\dfrac{\Delta {\cal S}}{k_B} = \zeta + T \dv{\zeta}{T} \ .
\end{equation}
To explore the physical intuition behind the results, let us refer to the resulting plots in detail. 
Figure (\ref{Entropy}) shows the dependence of $\Delta {\cal S} /k_B$ upon the normalized temperature $T/T_0$ for different values of relative polarization parameter  $\varsigma$. Increasing the $+$-mode relative to the $\times$-mode enhances the memory effect, consistent with the solutions discussed in section \ref{Sec:Pert.}, i.e. higher $\varsigma$ leads to a higher entropy change.

\begin{figure}[H]
\centering
\includegraphics[width=0.75\textwidth]{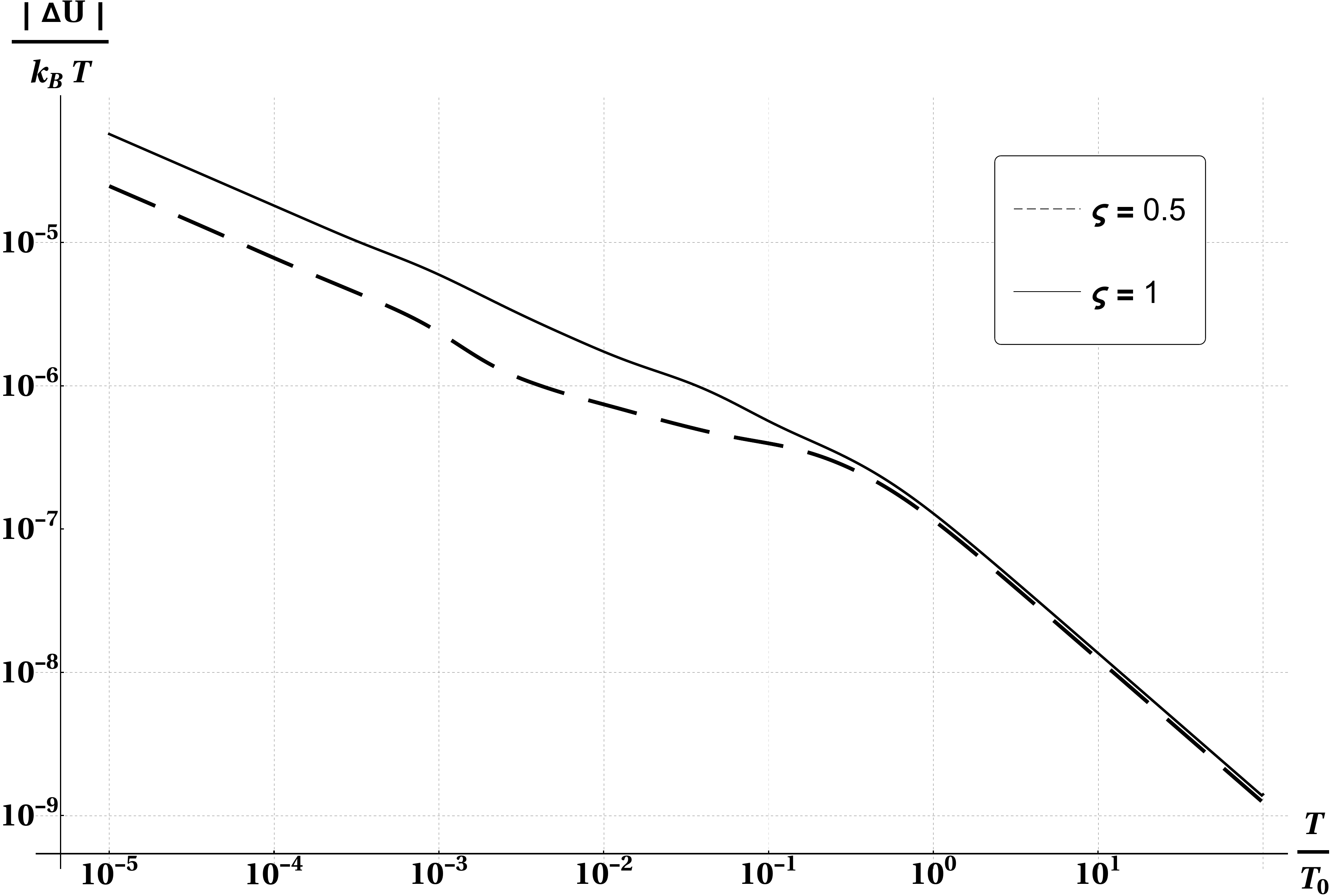} 
\caption{Change in the internal energy for small values of $\varsigma$.}
\label{Energy2}
\end{figure}

\begin{figure}[H]
\centering
\includegraphics[width=0.75\textwidth]{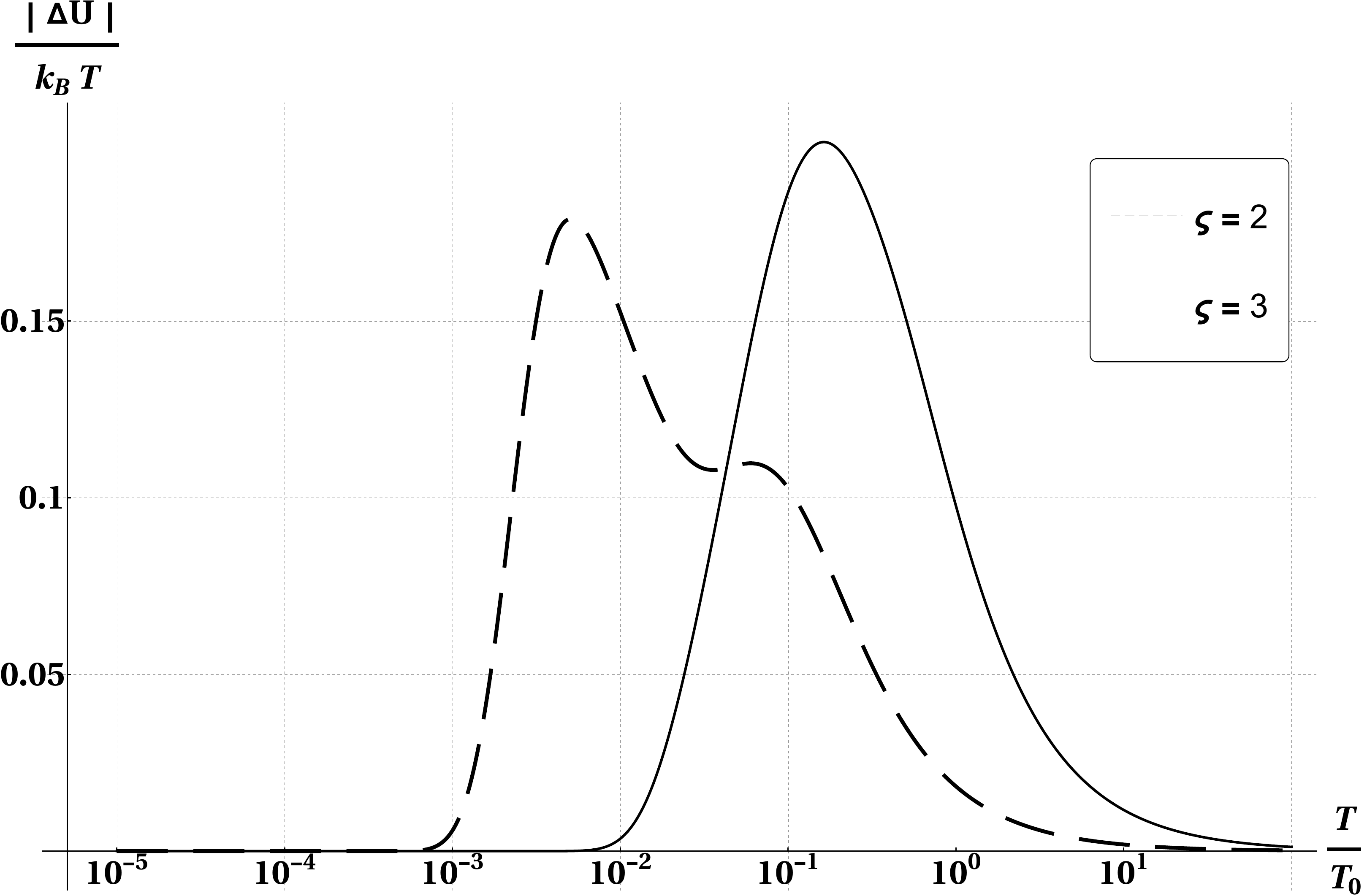} 
\caption{Change in the internal energy for large values of $\varsigma$.}
\label{Energy1}
\end{figure}

Another important thermodynamic quantity suitable for observing the hysteresis of the system is the internal energy, which reveals new aspects of the model. For $\varsigma = 0.5$ and $1$, results are as expected. As the temperature increases, the effectiveness of the memory sector of the partition function, $\Delta \mathcal{Z}$, decreases, meaning that the memory no longer significantly influences the relationship between entropy and energy.  Therefore, as expected, the energy follows the same pattern as the entropy, as shown in figure (\ref{Energy2}).
However, this pattern changes for large values of $\varsigma$. For $\varsigma = 2$ and $3$, some resonance behavior happens, as it is shown in figure (\ref{Energy1}). Such a behavior is quite natural. The system is able to absorb energy from the pulse at some temperature. With increasing the value of $\varsigma$ more alignment appears and the resonance temperature approaches the natural temperature of the system, $T_0$. 

Note that there is nothing special about the large values of $\varsigma$, i.e. waves essentially being $+$-type. The difference between $+$- and $\times$-types is oscillation along $x-y$ directions and $45^{\circ}$ directions in $x-y$ plane. As we initialized all the spins with $\phi=0$, we have chosen a preferred direction ($x$-axis). Therefore the more $+$-type, the more energy capture.

%%%%%%%%%%%%%%%%%%%%%%%%%%%%%%%%%%%% Section VI
\section{Concluding Remarks}
\label{Sec:Con.}

The non-linearity of Einstein equation is one of the distinctive features of general relativity, giving rise to hysteresis effects both in the systems interacting with gravity and in the gravitational field. This classical phenomenon, known in the literature as the memory effect, also manifests in the weak field regime, which is of special significance in the context of gravitational waves.

In this work, we examined how the dynamics of an extended rotating object can reveal traces of this memory effect. The Mathisson--Papapetrou--Dixon formalism describes extended body dynamics through the local interaction between the object's energy--momentum tensor and the background spacetime. Any perturbation in the spacetime acts as a source of nonhomogeneity in the equations of motion, as governed by equations \eqref{MP-Momentum} to \eqref{MP-Velocity}.

Using a perturbative analysis, we demonstrated that after the passage of a gravitational wave, the initial orientation of the object in the plane perpendicular to the direction of propagation exhibits a memory effect. This result is in full agreement with our previous analytical study in \cite{Moti:2024a}, which considered point particles rather than extended objects. Same happens for the linear motion of the object. The footprint of the passage of the gravitational wave would be encoded in both the linear motion and in the spin of any extended body.

According to general relativity, if the dynamics of the extended body is altered, it would radiates additional gravitational radiation. That is to say, the incoming wave would experiences some scattering. Therefore there would be (non-linear) memory effect in the gravitational wave, too. This effect was subsequently investigated. 

These considerations show that there is a possible way of knowing about the passage of a wave in the past through the body by examining its motion. Also the presence of a body on the way of a gravitational wave can lately be observed via the change in the wave. We saw that the change in orientation of the rotating body has a highly oscillatory nature with respect to the initial orientation. This can make difficult the observation of the effect. Using an ensemble of such objects can lead to a smooth behavior.
 
Considering an ensemble of rotating extended objects we show how the memory effect appears at the thermodynamic description of the system, as any redistribution of the spin of the elements of the ensemble, has footprints in the partition function. The partition function undergoes a change $\Delta\mathcal{Z}$ after interacting with the incoming wave. This, leads to deviation of statistical quantities like entropy and energy from their initial value, figures (\ref{Entropy}) to (\ref{Energy1}). We also observed that for properly chosen parameters, there is some kind of resonance at a specific temperature in the energy of the system, figure (\ref{Energy1}). 

As a result, here we demonstrated how the memory effect manifests in the thermodynamics of the system. Redistribution of spin orientations caused by the passing gravitational wave leaves its imprint on the partition function. This in turn leads to deviations of statistical quantities, reflecting the tendency of the ensemble to absorb energy from the wave.

%%%%%%%%%%%%%%%%%%%%%%%%%%%%%%%%%%%% References


\begin{thebibliography}{10}

\bibitem{Zeldovich:1974gvh}
Y.B. Zel'dovich and A.G. Polnarev, \textit{Radiation of gravitational waves by a cluster of superdense stars}, \href{https://ui.adsabs.harvard.edu/abs/1974AZh....51...30Z}{Sov. Astron. \textbf{18}  (1974) 17}.

\bibitem{Braginsky:1987kwo}
V.B. Braginsky and K.S. Thorne, \textit{Gravitational-wave bursts with memory and experimental prospects}, \href{https://doi.org/10.1038/327123a0}{Nature \textbf{327} (1987) 123}.

\bibitem{Christodoulou:1991cr}
D. Christodoulou, \textit{Nonlinear nature of gravitation and gravitational wave experiments}, \href{https://doi.org/10.1103/PhysRevLett.67.1486}{Phys. Rev. Lett. \textbf{67} (1991) 1486}.

\bibitem{Blanchet:1992br}
L. Blanchet and T. Damour, \textit{Hereditary effects in gravitational radiation}, \href{https://doi.org/10.1103/PhysRevD.46.4304}{Phys. Rev. D \textbf{46} (1992) 4304}.

\bibitem{Thorne:1992sdb}
K. S. Thorne, \textit{Gravitational-wave bursts with memory: The Christodoulou effect}, \href{https://doi.org/10.1103/PhysRevD.45.520}{Phys. Rev. D \textbf{45} (1992) 520}.

\bibitem{Divakarla:2021xrd}
A.K. Divakarla and B.F. Whiting, \textit{First-order velocity memory effect from compact binary coalescing sources}, \href{https://doi.org/10.1103/PhysRevD.104.064001}{Phys. Rev. D \textbf{104} (2021) 064001} [\href{https://arxiv.org/abs/2106.05163}{arXiv:2106.05163}].

\bibitem{Grishchuk:1989qa}
L.P. Grishchuk and A.G. Polnarev, \textit{Gravitational wave pulses with `velocity coded memory'}, \href{https://ui.adsabs.harvard.edu/abs/1989ZhETF..96.1153G/abstract}{Sov. Phys. JETP \textbf{69} (1989) 653}.

\bibitem{Favata:2009ii}
M. Favata, \textit{Nonlinear gravitational-wave memory from binary black hole mergers}, 
\href{https://doi.org/10.1088/0004-637X/696/2/L159}{Astrophys. J. Lett. \textbf{696} (2009)} [\href{https://arxiv.org/abs/0902.3660}{arXiv:0902.3660}].

\bibitem{Favata:2008ti}
M. Favata, \textit{Gravitational-wave memory revisited: Memory from the merger and recoil of binary black holes}, \href{https://doi.org/10.1088/1742-6596/154/1/012043}{J. Phys. Conf. Ser. \textbf{154} (2009) 012043} [\href{https://arxiv.org/abs/0811.3451}{arXiv:0811.3451}].

\bibitem{Favata:2008yd}
M. Favata, \textit{Post-Newtonian corrections to the gravitational-wave memory for quasi-circular, inspiralling compact binaries}, \href{https://doi.org/10.1103/PhysRevD.80.024002}{Phys. Rev. D \textbf{80} (2009) 024002} [\href{https://arxiv.org/abs/0812.0069}{arXiv:0812.0069}].

\bibitem{Favata:2010zu}
M. Favata, \textit{The gravitational-wave memory effect}, \href{https://doi.org/10.1088/0264-9381/27/8/084036}{Class. Quant. Grav. \textbf{27} (2010) 084036} [\href{https://arxiv.org/abs/1003.3486}{arXiv:1003.3486}].

\bibitem{Nichols:2017rqr}
D.A. Nichols, \textit{Spin memory effect for compact binaries in the post-Newtonian approximation}, \href{https://doi.org/10.1103/PhysRevD.95.084048}{Phys. Rev. D \textbf{95} (2017) 084048} [\href{https://arxiv.org/abs/1702.03300}{arXiv:1702.03300}].

\bibitem{Strominger:2014}
A. Strominger and A. Zhiboedov, \textit{Gravitational memory, BMS supertranslations and soft theorems}, \href{https://doi.org/10.1007/JHEP01(2016)086}{JHEP \textbf{01} (2014) 08} [\href{https://arxiv.org/abs/1411.5745}{arXiv:1411.5745}].

\bibitem{Pasterski:2015tva}
S. Pasterski, A. Strominger, and A. Zhiboedov, \textit{New gravitational memories}, \href{https://doi.org/10.1007/JHEP12(2016)053}{JHEP \textbf{12} (2016) 053} [\href{https://arxiv.org/abs/1502.06120}{arXiv:1502.06120}].

\bibitem{Zhang:2017geq}
P.M. Zhang, C.Duval, G.W. Gibbons, and P.A. Horvathy, \textit{Soft gravitons and the memory effect for plane gravitational waves}, \href{https://doi.org/10.1103/PhysRevD.96.064013}{Phys. Rev. D \textbf{96} (2017) 064013} [\href{https://arxiv.org/abs/1705.01378}{arXiv:1705.01378}].

\bibitem{Pate:2017}
M. Pate, A Raclariu, and A. Strominger, \textit{Color memory: A Yang-Mills analog of gravitational wave memory}, \href{https://doi.org/10.1103/PhysRevLett.119.261602}{Phys. Rev. Lett. \textbf{119} (2017) 261602} [\href{https://arxiv.org/abs/1707.08016}{arXiv:1707.08016}].

\bibitem{Luca:2025}
V.D. Luca, J. Khoury, and S.S.C. Wong, \textit{Gravitational memory and soft theorems: The local perspective}, \href{https://doi.org/10.1103/gbg1-mz49}{Phys. Rev. D \textbf{112} (2025) 2} [\href{https://arxiv.org/abs/2412.01910}{arXiv:2412.01910}].

\bibitem{Caldwell:2025}
R.R. Caldwell, \textit{The persistence of nonlinear gravitational wave memory}, [\href{https://arxiv.org/abs/2506.20751}{arXiv:2506.20751}].

\bibitem{Moti:2024a}
R. Moti and A. Shojai, \textit{On the gravitational precession memory effect for an ensemble of gyroscopes}, \href{https://doi.org/10.1088/1361-6382/ad1780}{Class. Quant. Grav. \textbf{41} (2024) 025011} [\href{https://arxiv.org/abs/2307.04151}{arXiv:2307.04151}].

\bibitem{Moti:2024b}
R. Moti and A. Shojai, \textit{On the gravitational hysteresis in the kinetic theory}, \href{https://doi.org/10.1088/1475-7516/2024/12/002}{JCAP \textbf{12} (2024) 002} [\href{https://arxiv.org/abs/2410.04537}{arXiv:2410.04537}].

\bibitem{Mathisson:1937}
M. Mathisson, \textit{Republication of: New mechanics of material systems}, \href{https://doi.org/10.1007/s10714-010-0939-y}{Gen. Rel. Grav. \textbf{42} (2010) 1011}.

\bibitem{Papapetrou:1951}
A. Papapetrou, \textit{Spinning test-particles in general relativity \Romannum{1}}, \href{https://doi.org/10.1098/rspa.1951.0200}{Proc. R. Soc. Lond. A \textbf{209} (1951) 248}.

\bibitem{Corinaldesi:1951}
E. Corinaldesi and A. Papapetrou, \textit{Spinning test-particles in general relativity \Romannum{2}}, \href{https://doi.org/10.1098/rspa.1951.0201}{Proc. R. Soc. Lond. A \textbf{209} (1951) 259}.

\bibitem{Dixon:1970}
W.G. Dixon, \textit{Dynamics of extended bodies in general relativity. \Romannum{1}. Momentum and angular momentum}, \href{https://doi.org/10.1098/rspa.1970.0020}{Proc. R. Soc. Lond. A \textbf{314} (1970) 499}.

\bibitem{Tod:1976}
K.P. Tod, F. De Felice, and M. Calvani, \textit{Spinning test particles in the field of a black hole}, \href{https://doi.org/10.1007/BF02728614}{Nuovo Cim. B \textbf{34} (1976) 365}.

\bibitem{Mashhoon:2006}
B. Mashhoon and D. Singh, \textit{Dynamics of extended spinning masses in a gravitational field}, \href{https://doi.org/10.1103/PhysRevD.74.124006}{Phys. Rev. D \textbf{74} (2006) 124006} [\href{https://arxiv.org/abs/astro-ph/0608278}{arXiv:astro-ph/0608278}].

\bibitem{Singh:2008}
D. Singh, \textit{An analytic perturbation approach for classical spinning particle dynamics}, \href{https://doi.org/10.1007/s10714-007-0597-x}{Gen. Rel. Grav. \textbf{40} (2008) 1179} [\href{https://arxiv.org/abs/0706.0928}{arXiv:0706.0928}].

\bibitem{Mohseni:2000}
M. Mohseni, and H.R. Sepangi, \textit{Gravitational waves and spinning test particles}, \href{https://doi.org/10.1088/0264-9381/17/22/302}{Class. Quant. Grav. \textbf{17} (2000) 4615}
[\href{https://arxiv.org/abs/gr-qc/0009070}{arXiv:gr-qc/0009070}].

\bibitem{Maggiore:2007ulw}
M. Maggiore, \textit{Gravitational Waves. Vol. 1: Theory and Experiments}, Oxford University Press (2007) [\href{https://doi.org/10.1093/acprof:oso/9780198570745.001.0001}{DOI: 10.1093/acprof:oso/9780198570745.001.0001}].


\end{thebibliography}
\end{document}